\documentclass[review,11pt]{elsarticle}

\journal{a journal}

\usepackage{amssymb}
\usepackage{amsmath}
\usepackage{setspace}
\usepackage{bm,upgreek}
\usepackage{xcolor,soul}
\usepackage{booktabs}
\usepackage{dirtytalk}

\makeatletter
\def\ps@pprintTitle{%
 \let\@oddhead\@empty
 \let\@evenhead\@empty
 \def\@oddfoot{\centerline{\thepage}}%
 \let\@evenfoot\@oddfoot}
 \makeatother

\usepackage{etoolbox}
\patchcmd{\MaketitleBox}{\footnotesize\itshape\elsaddress\par\vskip36pt}{\footnotesize\itshape\elsaddress\par\parbox[b][36pt]{\linewidth}{\vfill\hfill\textnormal{February 2024}\hfill\null\vfill}}{}{}%
\patchcmd{\pprintMaketitle}{\footnotesize\itshape\elsaddress\par\vskip36pt}{\footnotesize\itshape\elsaddress\par\parbox[b][36pt]{\linewidth}{\vfill\hfill\textnormal{February 2024}\hfill\null\vfill}}{}{}%

\usepackage[margin=1in]{geometry}
\usepackage{lineno}
\usepackage{hyperref} 
\hypersetup{hidelinks}

\selectfont

\definecolor{changes}{rgb}{1,0,0}

\newdefinition{axiom}{Axiom}
\newdefinition{definition}{Definition}
\newproof{proof}{Proof}

\biboptions{authoryear}

\begin{document}

\begin{frontmatter}

\title{Stochastic Nonparametric Estimation of the \textcolor{black}{Density-Flow Curve}}

\author[iaroslav]{Iaroslav Kriuchkov\corref{mycorrespondingauthor}}
\ead{iaroslav.kriuchkov@aalto.fi}
\author[timo]{Timo Kuosmanen}
\ead{timo.kuosmanen@utu.fi}

\cortext[mycorrespondingauthor]{Corresponding author}

\address[iaroslav]{Aalto University School of Business, 02150 Espoo, Finland}
\address[timo]{Turku School of Economics, University of Turku, 20500 Turku, Finland}
\date{May 2023}

\begin{abstract}
Recent advances in operations research and machine learning have revived interest in solving complex real-world, large-size traffic control problems. With the increasing availability of road sensor data, deterministic parametric models have proved inadequate in describing the variability of real-world data, especially in congested area of the density-flow diagram. In this paper we estimate the stochastic density-flow relationship introducing a nonparametric method called \textit{convex quantile regression}. The proposed method does not depend on any prior functional form assumptions, but thanks to the concavity constraints, the estimated function satisfies the theoretical properties of the \textcolor{black}{density-flow curve}. The second contribution is to develop the new \textit{convex quantile regression with bags} (CQRb) approach to facilitate practical implementation of CQR to the real-world data. We illustrate the CQRb estimation process using the road sensor data from Finland in years 2016-2018. Our third contribution is to demonstrate the excellent out-of-sample predictive power of the proposed CQRb method in comparison to the standard parametric deterministic approach.

\end{abstract}


\end{frontmatter}

\newpage
\section{Introduction} \label{sec:introduction}

Traffic congestion arises when the demand for the road space exceeds the available capacity. Congestion has numerous detrimental effects such as delays, increased fuel consumption, air pollution, vehicle wear and tear, and higher collision risks \citep{mirzaeian_queueing_2021, rus_constrained_2021}. Recognizing the negative consequences of congestion, researchers and policymakers have been exploring effective methods for managing traffic flow to alleviate congestion. One promising avenue that holds substantial potential lies in leveraging road sensor data, which provides valuable real-time insights into traffic conditions and patterns \citep{ shaygan_traffic_2022}. \par

Within the realm of the traffic flow theory, the fundamental diagram plays a pivotal role in understanding the relationship between traffic flow, density, and speed \citep{nagel_still_2003}. Originating from the seminal work of \citet{greenshields_study_1935} this concept has been extensively studied and applied in operations research. Traditionally, the density-flow relationship is modeled using simple parametric functional forms specifications such as piecewise linear triangular (bi-linear) forms \citep{newell_simplified_1993}, truncated piecewise linear triangular (bi-linear) forms \citep{daganzo_cell_1994}, or quadratic-linear forms \citep{smulders_control_1990}. Note that all these classic formulations assume a concave relationship between the density and flow. \par

In empirical research, deterministic approaches remain widely used for estimating \textcolor{black}{the density-flow or the speed-density relationship} from observed road sensor data \citep{lighthill_kinematic_1955, zhang_mathematical_1999, wang_logistic_2011, coifman_revisiting_2014}. However, these models are unable to predict the variance in the free-flow part of the \textcolor{black}{density-flow relationship}, let alone the congested part \citep{zhang_mathematical_1999}. Empirical road sensor data shows that such deterministic characterizations are far from reality due to the heterogeneity in types of vehicles, driver behavior, and the use of advanced driver-assistance systems (ADAS). To account for random heterogeneity in data, stochastic \textcolor{black}{density-flow curves} have started to gain popularity in the recent empirical literature \citep{hoogendoorn_new_1998, jabari_stochastic_2012, qu_stochastic_2017}.\footnote{Other attempts to describe heterogeneity include the models that allow for the capacity drop \citep{edie_car-following_1961}, three-phase approach \citep{kerner_structure_1994, kerner_introduction_2009} and accounting for differences between types of vehicles and drivers \citep{chanut_macroscopic_2003}.} Various approaches have been proposed to incorporate stochasticity into the fundamental diagram modeling. One such approach involves adding noise to deterministic relations, as demonstrated in works by \citet{muralidharan_probabilistic_2011} and \citet{wang_stochastic_2013}. Data-driven relation has gained momentum since then with numerous studies being conducted by \citet{fan_data-fitted_2013, qu_stochastic_2017} and \citet{wang_model_2021}. Despite these advances, it is important to note that these methods are still predominantly parametric, which means that they require a prior specification of the functional form. \par

This paper proposes a \textcolor{black}{stochastic,} fully nonparametric approach to estimating the \textcolor{black}{density-flow curve}.\footnote{The term nonparametric does not necessarily imply the absence of assumptions, and it is not necessarily the case that the assumptions of a nonparametric model are less restrictive than those of a parametric model. Following \citet{chen_large_2007}, an econometric model is termed “nonparametric” if all of its parameters are in infinite-dimensional parameter spaces.} Our method builds upon and extends the \textit{convex quantile regression} (CQR) approach by \citet{wang_nonparametric_2014, kuosmanen_shadow_2021} and \citet{dai_non-crossing_2023}, which belongs to the family of the convex regression models and related nonparametric estimators subject to convexity (concavity) constraints. To our knowledge, this study presents the first application of the convex regression approach to estimating the \textcolor{black}{density-flow curve} from empirical data. The proposed approach fits a piecewise linear function without any prior restrictions on the number of pieces or their location. We do not need to assume the piecewise linear functional form \textit{a priori}, rather, the piecewise linear functional form has been shown capable to represent any arbitrary concave function \citep{kuosmanen_representation_2008}.\footnote{The term convex regression refers to the fact that the support function of a convex set is globally concave or convex. Therefore, convex regression can be used for estimating globally concave or convex functions.} \textcolor{black}{To account for heterogeneity, multiple quantiles could be estimated (e.g. 0.7, 0.8, 0.9) and a distribution function for flow over density could be developed.} The proposed method satisfies all the theoretical properties of the \textcolor{black}{density-flow curve} noted by \citet{del_castillo_three_2012}. Quantiles have been independently used in the context of estimation the speed-flow relationship by \citet{wang_model_2021} however, they rely on the parametric specifications of the functional form whereas our approach is fully nonparametric. \par

Since the density-flow relationship is fundamentally concave, we would argue that the convex regression and CQR provide useful tools for the estimation in the stochastic setting with heterogeneous vehicle data. The concavity property was first recognized by \citet{ansorge_what_1990}, fully formulated by \citet{castillo_functional_1995-1} with its role explained within the variational theory of traffic flow by \citet{daganzo_variational_2005}. Although the concavity of the \textcolor{black}{density-flow curve} has also attracted some critical debate \citep{koshi_findings_1983, hall_freeway_1991}, we note that the solutions with deceleration shockwaves for the Lighthill–Whitham–Richards model \citep{lighthill_kinematic_1955, richards_shock_1956} can only be obtained in the concave case. \par

Our second contribution is to develop the new \textit{convex quantile regression with bags} (CQRb) method. CQRb allows computationally-intensive convex quantile regression to be applied to a large dataset. We illustrate the estimation process and the obtained function using the proposed method. The example is provided with the use of road sensor data from Finland (see Section \ref{sec:application} for details). \par

The third contribution is the first out-of-sample performance evaluation of the proposed method on the real-world road sensor data. Applying insights from the machine learning literature, we partition the observed sample to a training set and a test set to objectively assess the predictive power of the proposed method. In addition, we compare the performance of our method with the renowned method for the estimation of fundamental traffic flow parameters for triangular fundamental diagram introduced by \citet{dervisoglu_automatic_2009}. \par

The rest of the paper is organized as follows. In Section \ref{sec:FD} we discuss the estimation of the \textcolor{black}{density-flow curve} and introduce the idea of shape-constrained least-squares regression in additive formulation, further extending it to the nonparametric formulation with the introduction of quadratic programming problem. We further show how the aforementioned model is transformed into convex quantile regression with linear programming formulations in Section \ref{sec:quantiles}. After that we discuss the computational aspects of the estimations introducing bagging and the related convex quantile regression with bags (CQRb) in Section \ref{section:bagging}. We provide an illustrative application example to Finnish road sensor data in Section \ref{sec:application} and evaluate the performance of the proposed method in Section \ref{sec:performance}. Section \ref{sec:conclusion} concludes. \textcolor{black}{Additional comparisons and examples are presented in the appendix, which is available as an online supplement to this paper.}

\section{Preliminaries on the traffic flow theory and convex regression} \label{sec:FD}
\subsection{Fundamental relationship}

The fundamental diagram of traffic flow represents the relationship between three variables: traffic flow ($q$), traffic speed ($v$) and traffic density ($k$). The diagram can be projected onto three main planes: to density-flow plane $q=f(k)$, to density-speed plane and to speed-flow plane. These projections are interconnected through the fundamental relationship proposed by \citet{lighthill_kinematic_1955}:
\begin{align}\label{eq:fundrel}
    q = kv
\end{align}

There are two primary approaches to fundamental diagram estimation: via density-flow relationship \citep{smulders_control_1990, newell_simplified_1993, daganzo_cell_1994} and via density-speed relationship \citep{wu_new_2002, qu_stochastic_2017}. While both have been extensively studied in the literature, our focus is on the estimation of density-flow relationship $q=f(k)$, as the method we propose is inherently concave as discussed above. Moreover, it is widely accepted that increasing interaction between vehicles, meaning higher density, is the cause of the variance in traffic speed and traffic flow \citep{tadaki_critical_2015}. \par

In this method, we focus on concave shapes of density-flow relationship, as these forms are predominant in the traffic flow research with concavity property playing an important role in the literature \citep{ansorge_what_1990, castillo_functional_1995-1, daganzo_variational_2005}. It is worth mentioning that the concave shape of the \textcolor{black}{density-flow curve} is also recognized to be useful for traffic management purposes \citep{roncoli_optimal_2016}. The peak flow rate ($q_{c}$) and the corresponding density ($k_{c}$), as well as the jam density ($k_{j}$) can be easily derived from the concave \textcolor{black}{density-flow curve}. This way the strategies can be developed to minimize congestion  and to prevent the actual density from exceeding the critical density.

The density-flow relationship can be stated as a regression equation:\footnote{For clarity, we denote vectors with bold lowercase symbols. All vectors are column vectors. As the focus of this paper is density-flow relationship, the standard notation of \textit{k} for density and \textit{q} for flow is used.}
\begin{align}\label{eq:basicformulation}
    q_i = &f(\boldsymbol{\mathrm{k}}_i)+\varepsilon_i \quad \forall \, i, \: i=1,...,n
\end{align}
where $\boldsymbol{\mathrm{k}}_i = (k_{i1}, k_{i2}, k_{i3}, ..., k_{im})^\intercal$ represents a $m$-dimensional vector $\boldsymbol{\mathrm{k}} \in \mathbb{R}^m $ of predictors, including traffic flow $k$ (potentially observed on multiple lanes), and $q_i$ is a response variable for some observation $i$. The \textcolor{black}{random} error term $\varepsilon$ has a zero mean $\mathrm{E}(\varepsilon_i)=0 \; \; \forall i$ and finite variance $\mathrm{V}(\varepsilon_i)<\infty \; \; \forall i$. In the context of road traffic the error grasps the The \textcolor{black}{random} fluctuations in road conditions, for example weather and traffic composition. 

\subsection{Convex regression}
Suppose the regression function $f: \mathbb{R}^m \rightarrow \mathbb{R}$ is unknown but satisfies certain shape restrictions, such as concavity, monotonicity, and homogeneity (see \citet{kuosmanen_data_2010} and \citet{yagi_shape-constrained_2020} for a more detailed discussion). \citet{hildreth_point_1954} is the pioneer in the study of nonparametric regression with monotonicity and concavity constraints in the case of a single explanatory variable $k$. \citet{kuosmanen_representation_2008} further develops Hildreth's approach by extending it to the case of a vector-valued $\boldsymbol{\mathrm{k}}$ in a multivariate setting, and refers to this method as \textit{convex nonparametric least squares} (CNLS). In this CNLS method, it is assumed that the function $f$ belongs to a set of continuous, monotonically increasing, and globally concave functions called $F_2$, and $ \mathcal{F} \subset F_2$.  For the specifics of the \textcolor{black}{density-flow curve estimation}, we will focus only on the set of concave functions $\mathcal{F}=\{f:\mathbb{R}_+^m \rightarrow \mathbb{R}_+ \; \vert \;\forall \, \boldsymbol{\mathrm{k}}_1, \boldsymbol{\mathrm{k}}_2 \in \mathbb{R}_+^m : \boldsymbol{\mathrm{k}}_1 < \boldsymbol{\mathrm{k}}_2, \; f((1-\alpha)\boldsymbol{\mathrm{k}}_1 + \alpha \boldsymbol{\mathrm{k}}_2) \geq (1-\alpha)f(\boldsymbol{\mathrm{k}}_1) + \alpha f(\boldsymbol{\mathrm{k}}_2) \}$, which contains the regression function $f$. The CNLS estimator of function $f$ is defined as the optimal solution to the following infinite dimensional least squares problem:

\begin{align}\label{eq:infdimls}
     \min_{f} & \sum_{i=1}^n \upvarepsilon_i^2 \\
    \text{s.t.} \quad &  q_i = f(\boldsymbol{\mathrm{k}}_i) + \upvarepsilon_i \quad \forall i  \nonumber \\
    & f \in F_2 \nonumber
\end{align}
It is important to note that the functional form of $f$ is not predetermined. \textcolor{black}{To our knowledge, one cannot derive any specific functional form of the density-flow curve from physics.} Since $F_2$ consists of infinite number of functions, this problem cannot be solved using a simple trial-and-error approach. \par

\textcolor{black}{The representation theorem by \citet{kuosmanen_representation_2008} formally shows that the optimal solution $f^*$ to problem (\ref{eq:infdimls}) can always be equivalently characterized as a piece-wise linear function. Utilizing this result, we can harmlessly replace problem (\ref{eq:infdimls}) by the following finite dimensional quadratic programming problem:}
\begin{alignat}{2}\label{eq:additiveCNLS}
    \min_{\alpha, \bm{\upbeta}, \upvarepsilon} & \sum_{i=1}^n \upvarepsilon_i^2 \\
    \text{s.t.} \quad &  q_i = \alpha_i + \bm{\upbeta}_i^\intercal \boldsymbol{\mathrm{k}}_i + \upvarepsilon_i \quad &&\forall i  \nonumber \\
    & \alpha_i + \bm{\upbeta}_i^\intercal \boldsymbol{\mathrm{k}}_i \leq \alpha_s + \bm{\upbeta}_s^\intercal \boldsymbol{\mathrm{k}}_i \quad &&\forall i, s \quad i \neq s \nonumber
\end{alignat}

In problem (\ref{eq:additiveCNLS}), the objective function minimizes the sum of squared errors. Coefficients $\alpha_i$ and $\bm{\upbeta}_i$ are the intercept and slope coefficients of the tangent hyperplanes that describe the estimated piecewise linear function $\hat{f}$. The multiplication $\bm{\upbeta}_i' \boldsymbol{\mathrm{k}}_i$ is a vector product of a transposed vector of slope coefficients $\bm{\upbeta}_i$ and a vector of explanatory variables $\boldsymbol{\mathrm{k}}_i$ $(\bm{\upbeta}_i^\intercal \boldsymbol{\mathrm{k}}_i = \beta_{i1}k_{i1} + \beta_{i2}k_{i2} +...+\beta_{im}k_{im})$. The first constraint of the problem expresses the regression equation (\ref{eq:basicformulation}) in terms of a piecewise linear approximation of the true, yet unknown, regression function $f$. The second constraint requires the concavity of the estimated function (to impose convexity, the inequality sign should be reversed). In \citet{kuosmanen_representation_2008} the monotonicity assumption is imposed by adding the constraint $\bm{\upbeta}_i \geq \bm{0} \; \forall i$, but in the present context we relax it to allow the downward sloping part of the curve to exist.  \par 

The solution of problem (\ref{eq:additiveCNLS}) yields a unique pair of slope and intercept coefficients $(\hat{\alpha_i}, \bm{\hat{\upbeta}}_i)$ for each observation $i$, resulting in the best possible fit. The piecewise linear function $f$ is constructed using these pairs $(\hat{\alpha_i}, \bm{\hat{\upbeta}}_i)$, meaning $\hat{f}_i = \hat{\alpha_i} + \bm{\hat{\upbeta}}_i\boldsymbol{\mathrm{k}}_i$. In practice, the number of segments of a piecewise linear function is considerably smaller than the number of observations. In the estimation of density-flow relationship the meaning of $\upbeta_i$ is the shockwave speed. \par

\begin{figure}[h]
   \centering
   \includegraphics[width=\textwidth]{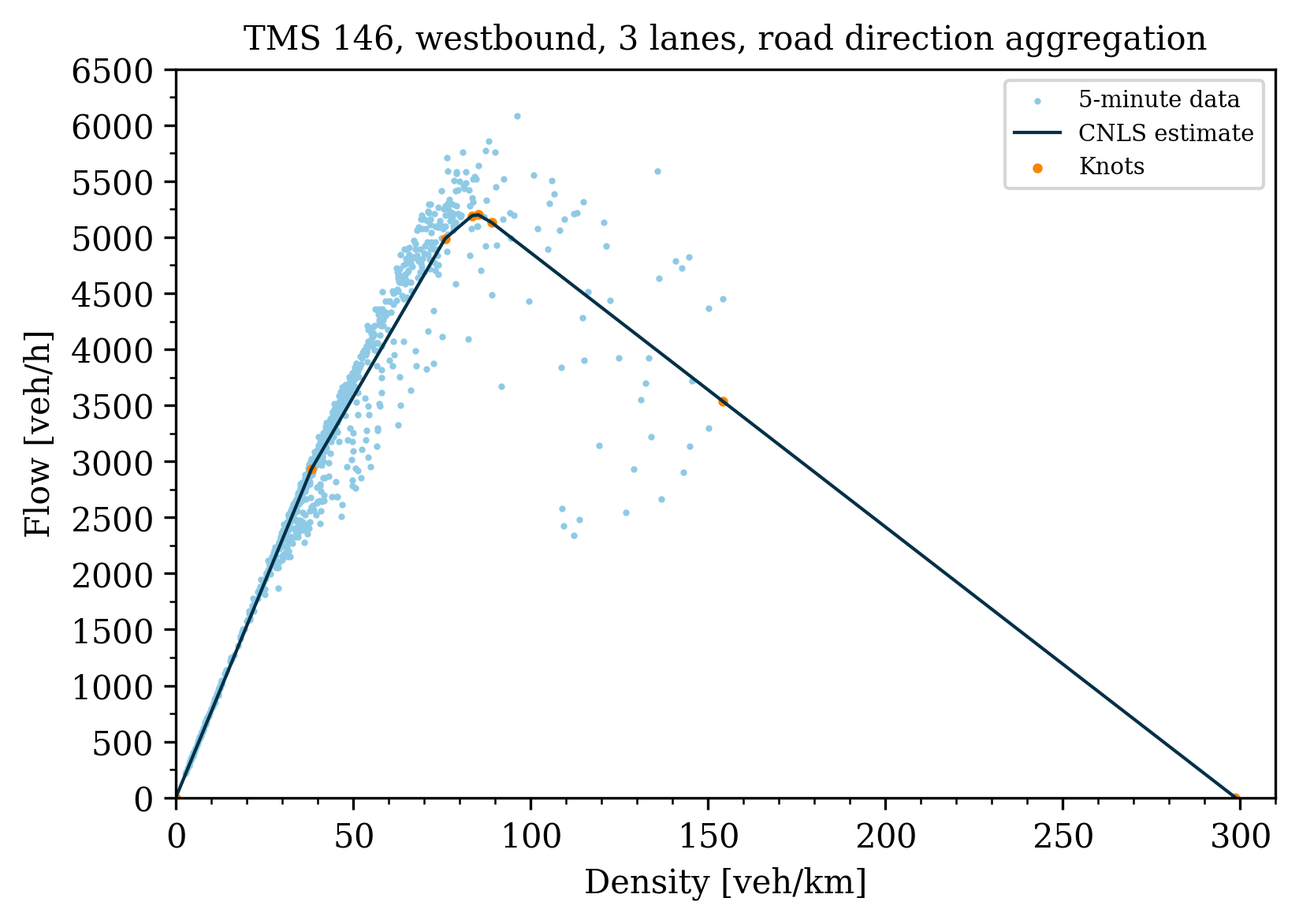}
   \caption{Density-flow curve estimated with CNLS}
   \label{fig:CNLS_example}
\end{figure}

Figure \ref{fig:CNLS_example} illustrates the example of the CNLS estimation on 5-minute-data from a traffic measurement station \textcolor{black}{with 3 lanes, and data being aggregated per road direction} (see Section \ref{sec:application} for a detailed description of the data). Knots on the figure illustrate the connections between different pieces. \textcolor{black}{The jam density is calculated as $\frac{-\hat{\alpha}_{\max \mathrm{k}}}{\hat{\upbeta}_{\max \mathrm{k}}}$.}  Note that the shockwave speed gradually decreases as density $k$ increases. At the critical density of $k=85.29$, the sign of the shockwave speed becomes negative, as illustrated by the downwards sloping part of the density-flow diagram in Figure \ref{fig:CNLS_example}. The estimated piecewise linear function depicted in Figure \ref{fig:CNLS_example} can be explicitly stated as follows. \par

\begin{align}\label{eq:CNLS_estimated_f}
\scriptsize
\hat{f}(k)=
\begin{cases}
    76.88k + 45.12    & 0       \leq k\leq 38.29 \\
    55.48k + 864.03   & 38.29   \leq k\leq 75.93 \\
    15.37k + 3910.85  & 75.93   \leq k\leq 83.54 \\
    2.93k + 4951.16   & 83.54   \leq k\leq 85.29 \\
    -16.91k + 6644.69 & 85.29   \leq k\leq 89.07 \\
    -24.34k + 7306.55 & 89.07   \leq k\leq 300.19 \\
\end{cases}
\end{align}%

For a univariate case $m=1$ of problem (\ref{eq:additiveCNLS}) the convexity constraint can be simplified following \citet{hanson_consistency_1976} and \citet{lee_more_2013} for more efficient computations. The simplification requires the sorting of the observed data in ascending order according to $k$. The simplified problem has the following formulation (\ref{eq:additiveCNLS_univariate}).

\begin{alignat}{2}\label{eq:additiveCNLS_univariate}
    \min_{\alpha, \upbeta, \upvarepsilon} & \sum_{i=1}^n \upvarepsilon_i^2 \\
    \text{s.t.} \quad &  q_i = \alpha_i + \upbeta_i k_i + \upvarepsilon_i \quad &&\forall i  \nonumber \\
    & \upbeta_i \leq \upbeta_{i-1}\quad &&\forall i, i = 2,...,n \nonumber \\
    & \alpha_i \geq \alpha_{i-1} \quad &&\forall i, i = 2,...,n \nonumber
\end{alignat}

The concept of using a piecewise linear function $\hat{f}$ is applied in a different context to the estimation of the macroscopic fundamental diagram (MFD). \citet{daganzo_analytical_2008} propose an analytical approximation of the MFD using a technique referred to as \say{tight cuts}, which is similar in spirit to the CNLS method. \par

\section{Convex quantile regression}\label{sec:quantiles}

Traffic data are known to be widely scattered \citep{treiber_understanding_2006}, and various explanations have been proposed for this phenomenon, including stochastic effects \citep{treiber_understanding_2006} and the heterogeneity of traffic \citep{knoop_automatic_2017}. The congested part of the \textcolor{black}{density-flow diagram} has substantially higher variance compared to the uncongested part. Furthermore, estimating the conditional mean flow for a given density is not always a priority for traffic researchers, as it requires a deterministic capacity value and may bias the shockwave speed. To address these issues, we propose to use the quantile regression to estimate the density-flow \textcolor{black}{curve}. \textcolor{black}{Estimation of multiple quantiles allows to account for the heterogeneity in traffic data. Different quantiles can reveal the spread of values in the congested region and identify potential extremes.}  Quantile regression, in one way or another, has been used to estimate the \textcolor{black}{density-flow or density-speed relationship} in the works of \citet{dervisoglu_automatic_2009}, \citet{li_fundamental_2011} and \citet{wang_model_2021}. We next expand the quantile regression to convex quantile regression (CQR), extending the CNLS approach. \par

Convex nonparametric least squares (CNLS) estimates the conditional mean $\mathrm{E} (q_i\vert\boldsymbol{\mathrm{k}}_i)$ of the response variable. Quantile regression serves as a generalization of classical mean regression model by estimating some conditional quantile $Q_{q}(\tau\vert\boldsymbol{\mathrm{k}}), \tau \in (0, 1)$ \citep{koenker_regression_1978, koenker_quantile_2005}:
\begin{align}
    Q_{q}(\tau\,\vert\,\boldsymbol{\mathrm{k}}) = F^{-1}(\tau\,\vert\,\boldsymbol{\mathrm{k}}) = \text{inf}\{q \geq 0 \; \vert \; F(q\,\vert\,\boldsymbol{\mathrm{k}})\geq\tau\}
\end{align}
where $F(q\,\vert\,\boldsymbol{\mathrm{k}})$ is the conditional distribution function of $q$ given $\boldsymbol{\mathrm{k}}\leq\mathrm{k}$. If the observed $\boldsymbol{\mathrm{k}}$ are exogenous, then the quantile function can be equivalently stated \citep{dai_generalized_2023} as
\begin{align}
    Q_{q}(\tau\,\vert\,\boldsymbol{\mathrm{k_i}}) = f(\boldsymbol{\mathrm{k_i}}) +F_{\varepsilon_i}^{-1}(\tau)
\end{align}

In the context of convex regression, the \textit{convex quantile regression} (CQR) provides an analogous extension to CNLS. In practice, what differentiates CQR from CNLS is the loss function. While CNLS uses the symmetric least squares as the objective function, CQR utilizes the asymmetric least absolute loss, also known as quantile loss function $\rho_\tau(\xi) = \xi(\tau-\mathbb{I}_{\xi<0})$, where $\mathbb{I}_A$ is an indicator function. Following \citet{wang_nonparametric_2014, kuosmanen_stochastic_2015} and \citet{dai_non-crossing_2023} we extend the CNLS problem to estimate CQR in the following section. \par

\subsection{Convex quantile regression (CQR) programming problem}
Denoting a pre-specified quantile as $\tau \in (0, 1)$, the \textit{convex quantile regression} (CQR) linear programming problem is formulated as:
\begin{alignat}{2}\label{eq:additiveCQR}
    \min_{\alpha, \bm{\upbeta}, \upvarepsilon^+, \upvarepsilon^-} & \tau\sum_{i=1}^n \upvarepsilon_i^+ + (1-\tau)\sum_{i=1}^n \upvarepsilon_i^- \\
    \text{s.t.} \quad & q_i = \alpha_i + \bm{\upbeta}_i^\intercal \boldsymbol{\mathrm{k}}_i + \upvarepsilon_i^+ - \upvarepsilon_i^- \quad &&\forall i  \nonumber \\
    & \alpha_i + \bm{\upbeta}_i^\intercal \boldsymbol{\mathrm{k}}_i \leq \alpha_s + \bm{\upbeta}_s^\intercal \boldsymbol{\mathrm{k}}_i \quad &&\forall i, s, \quad i \neq s \nonumber \\
    & \upvarepsilon_i^+ \geq 0, \; \upvarepsilon_i^- \geq 0 \quad &&\forall i \nonumber
\end{alignat}

Here the error term consists of two non-negative components $\upvarepsilon^+, \upvarepsilon^- \geq 0$, so the objective function minimizes the absolute deviations instead of the symmetric quadratic. The first three constraints of problem (\ref{eq:additiveCQR}) are the same as those of the CNLS problem (\ref{eq:additiveCNLS}). The last constraint makes the components of additive error term non-negative. Using the same approach as for problem (\ref{eq:additiveCNLS_univariate}), it is possible to formulate a simplified approach for a univariate case $m=1$.\par

The pre-specified quantile $\tau\in (0, 1)$ defines the quantile to be estimated. For example, by setting the $\tau=0.9$, the piecewise linear CQR function envelopes at most 90\% of observations, and at most 10\% of the observations lie above the fitted function. With $\tau=0.5$ the conditional median is estimated.  \par

One key difference between CNLS and CQR is that the objective function for CQR is linear. With all constraints being linear functions of unknown parameters, the CQR problem (\ref{eq:additiveCQR}) can be solved using standard linear programming (LP) algorithms. However, the simplicity of the current specification has its own drawback – when several quantiles are estimated, some of the quantile curves may cross each other. In the following subsection the \textit{penalized convex expectile regression} (pCQR) is introduced to solve this drawback. 

\subsection{Penalized convex quantile regression (pCQR)}
To solve the crossing quantile problem \citet{dai_non-crossing_2023} propose the following formulation of penalized convex quantile regression using the $L_2$-norm regularization on subgradients $\bm{\upbeta}_i$:
\begin{alignat}{2}\label{eq:pCQR}
    \min_{\alpha, \bm{\upbeta}, \upvarepsilon^+, \upvarepsilon^-} & \tau\sum_{i=1}^n \upvarepsilon_i^+ + (1-\tau)\sum_{i=1}^n \upvarepsilon_i^- + \gamma \sum_{i=1}^n \Vert \bm{\upbeta}_i \Vert_2^2 \\
    \text{s.t.} \quad & q_i = \alpha_i + \bm{\upbeta}_i^\intercal \boldsymbol{\mathrm{k}}_i + \upvarepsilon_i^+ - \upvarepsilon_i^- \quad &&\forall i  \nonumber \\
    & \alpha_i + \bm{\upbeta}_i^\intercal \boldsymbol{\mathrm{k}}_i \leq \alpha_s + \bm{\upbeta}_s^\intercal \boldsymbol{\mathrm{k}}_i \quad &&\forall i, s, \quad i \neq s \nonumber \\
    & \upvarepsilon_i^+ \geq 0, \; \upvarepsilon_i^- \geq 0 \quad &&\forall i \nonumber
\end{alignat}

Here the $\gamma \geq 0$ is the tuning parameter, and $\Vert \cdot \Vert_2^2$ is the standard Euclidean norm. With $\gamma \rightarrow 0$ the pCQR problem (\ref{eq:pCQR}) collapses to the original CQR problem (\ref{eq:additiveCQR}). \par 

The pCQR method avoids quantile crossing based on the rationale that as the regularization parameter $\gamma \rightarrow \inf$, the regularization term dominates the minimization process, causing all estimated subgradients $\bm{\upbeta}_i$ to converge towards $0$. This leads to the estimated quantile functions being represented by horizontal lines (in the 1-dimensional cases) or planes (in the multidimensional cases). It means, that with a large enough $\gamma$, the quantile crossing can clearly be avoided. \citet{dai_non-crossing_2023} propose to iteratively find the smallest $\gamma$ for which no quantile crossing occurs. In practice, the suitable value of $\gamma$ also depends on the selected quantiles $\tau$ to be estimated. Finally, an added advantage of the penalty term is the guaranteed uniqueness of the optimal $(\hat{\alpha_i}, \bm{\hat{\upbeta}}_i)$ coefficients.\par

\section{Bagging}\label{section:bagging}
The CNLS problem (\ref{eq:additiveCNLS}) and CQR problem (\ref{eq:additiveCQR}) can be solved using the state-of-the-art quadratic programming (QP) and linear programming (LP) solvers, respectively. While these solvers are effective for small datasets (about several hundreds of observations), they are not well-suited to large datasets, such as those commonly encountered in traffic data. For example, data from a 2-lane loop detector for just 10 days of peak hours can contain as many as 3360 observations. \textcolor{black}{As a result, the quadratic programming problem has over 11 million constraints and becomes computationally expensive to solve.} Furthermore, for the CQR problem on the density-flow plane adding one more observation will increase the number of unknown parameters by 4 and the number of convexity constraints by $2n$, where $n$ is an overall number of observations. Also, the data on the density-flow plane is skewed: in the free-flow part of the diagram there are numerous observations, that do not bring much information, compared to a relatively scarce, but important congested part of the diagram. \par

To facilitate the estimation of the \textcolor{black}{density-flow curve} on arbitrary large datasets, we propose a computational technique referred to as \textit{bagging}. There exist different approaches introduced by, for example, \citet{hannah_multivariate_2013} and \citet{qu_stochastic_2017}, however, we develop a tailored approach for the convex quantile regression. \par 

In practice, \textit{bagging} involves restricting the number of observations to a level at which linear programming solvers can perform efficiently with the minimal loss of precision. This is achieved by aggregating the data on the density-flow plane and dividing it into a set of \textit{bags}, each of which is represented by a centroid describing the data points contained within the bag. The number of bags is determined by dividing the $k$-axis into $u$ segments and the $q$-axis into $v$ segments, with the intersection of each $u$ and $v$ segment defining a bag. Each bag contains a set of points $S_j = \{(k_1, q_1), (k_2, q_2), ..., (k_p, q_p)\}, \; p = \vert S_j \vert, \; j \in U \times V, \; U = \{1,...,u\}, V = \{1,...,v\}$, and the centroid, a representative point $(\mathrm{k}_j^{bag}, \mathrm{q}_j^{bag})$ for a bag $j \in U \times V$, is calculated as:
\begin{align}\label{eq:baggingformula}
  (\mathrm{k}_j^{bag}, \mathrm{q}_j^{bag}), \; \mathrm{k}_j^{bag} = \sum_{i=1}^p \frac{k_i}{p}, \; \mathrm{q}_j^{bag} = \sum_{i=1}^p \frac{q_i}{p}
\end{align} \par

As bags may contain varying numbers of observations, it is necessary to account for this discrepancy. To do so, we calculate the weight of each bag based on the number of observations it contains. Specifically, the weight of a bag $w_j = p_j/n$ is calculated as the ratio of the number of observations contained within the bag $p_j$ to the total number of observations $n$. Therefore, $0 \leq w_j \leq 1  \; \forall j \in \vert U \times V \vert$ and $\sum_{j=1}^{\vert U \times V \vert} w_j = 1$. \par

\begin{figure}[h]
    \centering
    \includegraphics[scale=0.8]{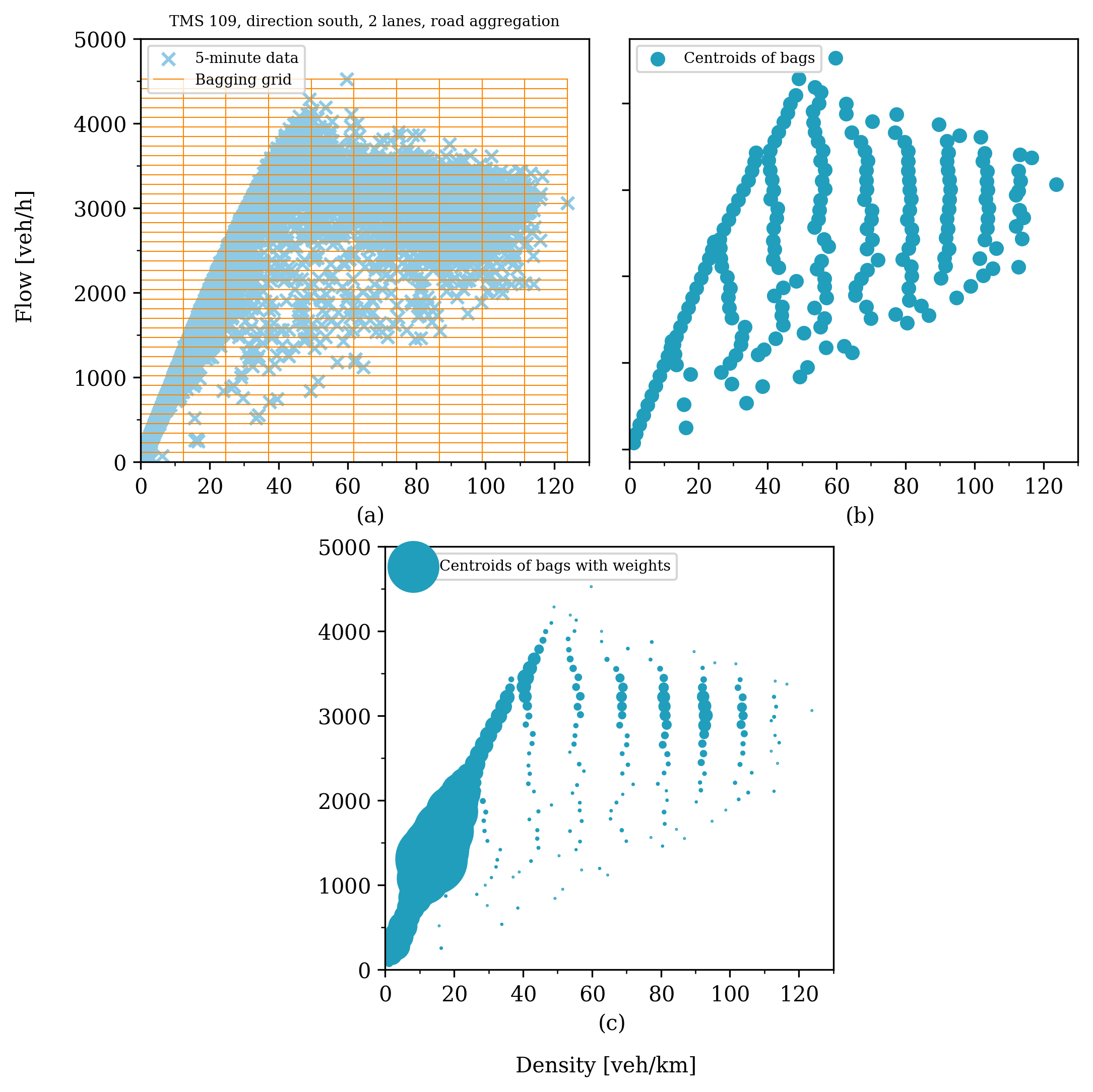}
    \caption{Visualisation of bagging process}
    \label{fig:bagging}
\end{figure}

Figure \ref{fig:bagging} illustrates an example of the bagging process applied to a dataset of 5-minute yearly observations from a traffic measurement station in Finland \textcolor{black}{with 2 lanes} over a period of one year, comprising 61109 observations \textcolor{black}{aggregated by road direction}. In this example, the bagging grid has a relatively small number of bags ($u=10$, $v=40$) for the purpose of demonstrating the approach. As shown in part b) of Figure \ref{fig:bagging}, the application of the bagging process results in a substantial reduction in the number of observations, which are to be represented by the centroids of the bags. However, it is important to note that the quantiles in this case would differ from the original dataset, as centroids do not account for the varying density of observations within the original dataset. To address this issue, the bagging process also assigns weights to the bags, as depicted on the bottom chart, where the size of each dot represents the weight of the corresponding bag. The higher the weight, the larger the size of the dot. This representation shows that the bagging method takes into account the density of the original observations, ensuring that unique or densely congested parts of the dataset are not overlooked. \textcolor{black}{Note that bagging is robust to outliers because potential outliers end up in bags with very few observations which carry a negligibly small weight in the estimation.} \par

The CQR problem, that is formulated for the representative points of the bags (\ref{eq:baggingformula}) and accounts for the weights of the bags, is referred to as \textit{convex quantile regression with bags} (CQRb) and has the following mathematical formulation: \\
\begin{alignat}{2}\label{eq:weightedtrafficCQR}
    \min_{\alpha, \upbeta, \upvarepsilon^+, \upvarepsilon^-} & \tau\sum_{j=1}^{\vert U \times V \vert} w_j\upvarepsilon_j^+ + (1-\tau)\sum_{j=1}^{\vert U \times V \vert} w_j\upvarepsilon_j^- + \gamma \sum_{j=1}^{\vert U \times V \vert} \Vert \bm{\upbeta}_j \Vert_2^2 \\
    \text{s.t.} \quad & \mathrm{q}_j^{bag} = \alpha_j + \upbeta_j \mathrm{k}_j^{bag} + \upvarepsilon_j^+ - \upvarepsilon_j^- \quad &&\forall j \in \vert U \times V \vert  \nonumber \\
    & \alpha_j + \upbeta_j \mathrm{k}_j^{bag} \leq \alpha_h + \upbeta_h \mathrm{k}_j^{bag} \quad &&\forall j, h \in \vert U \times V \vert, \quad j \neq h \nonumber \\
    & \upvarepsilon_j^+ \geq 0, \; \upvarepsilon_j^- \geq 0 \quad &&\forall j \in \vert U \times V \vert \nonumber
\end{alignat}
In the CQRb problem, the weights $w_j$ are applied to the errors such that the errors of bags with a larger number of observations are given more importance. The estimation is performed using the representative points $(\mathrm{k}_j^{bag}, \mathrm{q}_j^{bag})$ for each bag. In comparison to the classic CQR problem (\ref{eq:additiveCQR}), the CQRb problem does not have a monotonicity constraint and includes additional constraints on the weights. The jam density can be obtained as $\frac{-\hat{\alpha}_{\max \mathrm{k}}^{bag}}{\hat{\upbeta}_{\max \mathrm{k}}^{bag}}$. The solution of problem (\ref{eq:weightedtrafficCQR}) yields a unique pair of slope and intercept coefficients $(\hat{\alpha_j}, \hat{\upbeta}_j)$for each bag $j$, resulting in the best possible fit. The piecewise linear function $q$ is constructed using these pairs $(\hat{\alpha_j}, \hat{\upbeta}_j)$, meaning $\hat{q}_j^{bag} = \hat{\alpha_j} + \hat{\upbeta}_j\mathrm{k}_j^{bag}$. In practice, the number of segments of a piecewise linear function is substantially smaller than the number of observations. In the estimation of density-flow relationship the meaning $\upbeta_j$ is the shockwave speed. To satisfy the property, that at zero density there is zero flow the respective point should be added into the dataset for the estimation. \par

\section{Application}\label{sec:application}

\begin{figure}[ht]
    \centering
    \includegraphics[width=\textwidth]{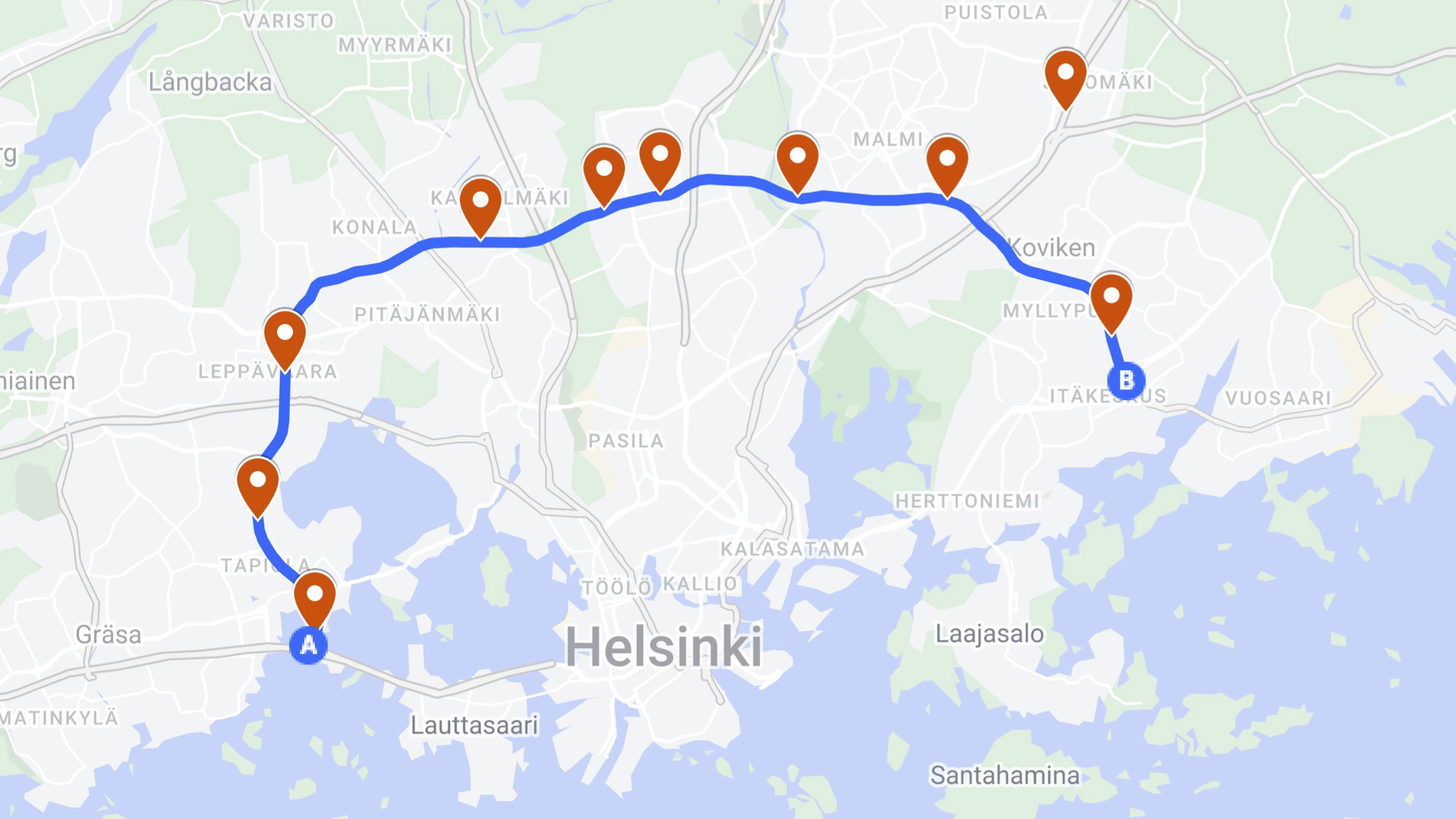}
    \caption{Location of traffic measurement stations around Ring I in Helsinki}
    \label{fig:tms_map}
\end{figure}

For the empirical research, we utilize traffic data provided by \citet{fintraffic_information_2023} under the Creative Commons 4.0 BY license. There are over 450 traffic measurement stations (TMS) in Finland, but for the purposes of this research, we focus on those located at the busiest road in the Capital region (which includes Helsinki, Espoo, Vantaa, and Kauniainen) and Finland overall, the Ring I (Finnish: Kehä I), numbered Regional Road 101. This 24-kilometer beltway allows vehicles to bypass the center of the capital of Finland, connecting Espoo with the eastern part of Helsinki. Figure \ref{fig:tms_map} shows the location of the traffic measurement stations (marked in red) on Ring I (marked blue). Notably, one station is located out of Ring I before the intersection of the Finnish national roads 4 (Lahdenväylä) and 7 (Porvoonväylä) and Ring I, and is of interest due to the high volume of traffic passing through. The numbers of traffic measurement stations, referred to in the following sections, from A to B on the map are 118, 10, 116, 145, 146, 147, 148, 149, 11, and the one not on Ring I is 109. \par

\textcolor{black}{Our data spans a 3-year period from January 1, 2016 to December 31, 2018. For every day a separate file containing raw data about all the vehicles passing the given loop detector is acquired. This file contains a special \texttt{faulty} flag, which is set to 1 if the observation has some incorrect parameters (time, speed, lanes, length, etc.). The data are cleaned based on this flag. There has been reduced availability of data from detectors 10 and 11, which are available only from early 2018. Also, detector 148 does not have information prior to autumn 2016. On average, there is from 0.2\% to 2.3\% of faulty observations per day depending on the detector. The number of vehicles observed is between 30 000 and 90 000 per day both directions combined. More extensive information on every detector is provided in Table \ref{tab:data-info}.} \par

\begin{table}[ht]
\centering
\caption{Summary of traffic measurement stations}
\label{tab:data-info}
\resizebox{\textwidth}{!}{%
\begin{tabular}{@{}lrrrrrrrrrr@{}}
\toprule
\textbf{Traffic Measurement Station} & \multicolumn{1}{c}{\textbf{10}} & \multicolumn{1}{c}{\textbf{11}} & \multicolumn{1}{c}{\textbf{109}} & \multicolumn{1}{c}{\textbf{116}} & \multicolumn{1}{c}{\textbf{118}} & \multicolumn{1}{c}{\textbf{145}} & \multicolumn{1}{c}{\textbf{146}} & \multicolumn{1}{c}{\textbf{147}} & \multicolumn{1}{c}{\textbf{148}} & \multicolumn{1}{c}{\textbf{149}} \\ \midrule
Data availability (\% of days)       & 28,4 \%                         & 29,7 \%                         & 99,7 \%                          & 96,4 \%                          & 99,7 \%                          & 97,5 \%                          & 91,6 \%                          & 99,2 \%                          & 75,7 \%                          & 98,3 \%                          \\
Average daily flow (vehicles)        & 37209                           & 47115                           & 49817                            & 60308                            & 30655                            & 82844                            & 93691                            & 91384                            & 60795                            & 62472                            \\
\% of false observations per day     & 0,52 \%                         & 0,45 \%                         & 0,43 \%                          & 0,24 \%                          & 1,02 \%                          & 1,89 \%                          & 1,18 \%                          & 0,22 \%                          & 0,06 \%                          & 2,34 \%                          \\ \bottomrule
\end{tabular}%
}
\end{table}

\textcolor{black}{In the following subsections application examples and comparisons are introduced. In Section \ref{sec:bagging_comparison} CQR and CQRb estimates of density-flow curves are compared as well as the computational burden of both methods. In Section \ref{sec:nearly_stationary} two approaches to data sampling are introduced: partitioning from nearly stationary data and time-aggregated data. The proposed method is applied on both and results are compared. Section \ref{sec:time_agg} provides an example of estimation using CQRb with all the estimated parameters directly shown. Finally, Section \ref{sec:overview} provides a comprehensive overview of estimates across different traffic measurement stations.} \par 

\subsection{\textcolor{black}{CQR versus CQRb comparison}} \label{sec:bagging_comparison}

To evaluate the burden of the CQR method (\ref{eq:additiveCQR}) and the CQRb method (\ref{eq:weightedtrafficCQR}) the pyStoNED package \citep{dai_pystoned_2024}  with local optimization is employed. The comparison is conducted on 5-minute data for 10 consecutive days, as using a larger number of days results in insufficient iterations to solve the problem on raw data, leading to termination of the solver. The results of the comparison are presented in Table \ref{tab:comparison_CQR_CQRb} and show the average performance of the two methods. \par

\begin{table}[h]
\centering
\small
\caption{Comparison of computational burden of CQR and CQRb}
\vspace{10pt}
\begin{tabular}{@{}lcc@{}}
\toprule
                      & CQR           & CQRb        \\ \midrule
Data points           & 3 360         & 643         \\
Number of constraints & 11 292 960    & 711 492     \\
Time elapsed          & 2 060 seconds & 112 seconds \\
\bottomrule
\end{tabular}
\label{tab:comparison_CQR_CQRb}
\end{table}

\begin{figure}[ht]
    \centering
    \includegraphics[width=\textwidth]{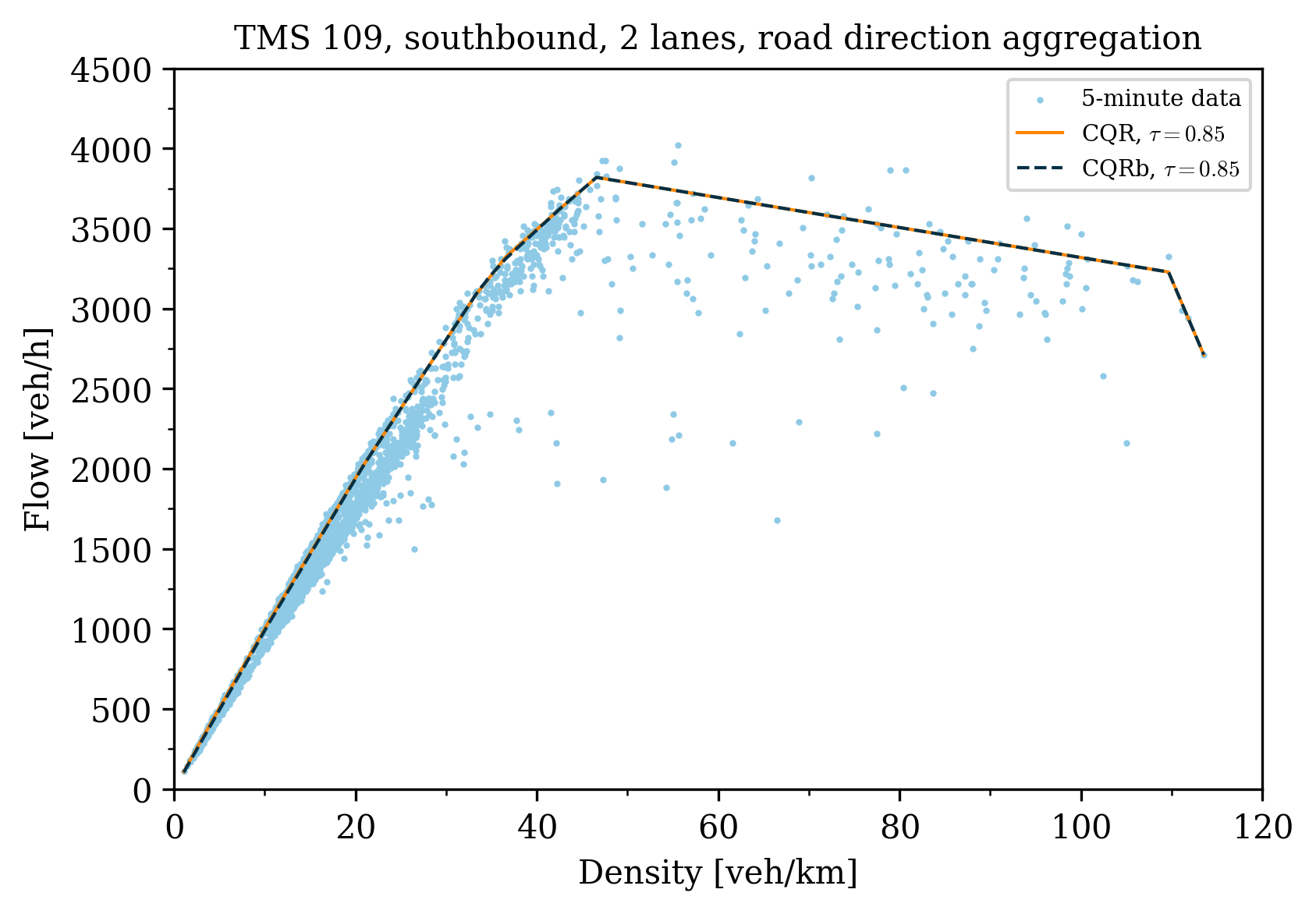}
    \caption{Comparison of CQR and CQRb}
    \label{fig:CQR_CQRb_example}
\end{figure}

An example of this is shown in Figure \ref{fig:CQR_CQRb_example} where the orange line represents the estimation on the original 5-minute data \textcolor{black}{from a 2-lane road} and the blue dashed line represents the extrapolation of the CQRb estimation from the bagged data to the original data. It can be observed that the estimated functions are almost identical in shape. While both CQR and CQRb are similar in terms of shape, they differ in computational complexity. \par 

It is observed that bagging results in a substantially smaller number of data points for CQRb compared to CQR, with the number of data points being more than five times smaller for CQRb. This leads to a reduction in the number of constraints by approximately 16 times. Additionally, the time required for the estimation is reduced from 34 minutes for CQR to less than 2 minutes for CQRb, including the time for extrapolation from the bagged data to the original 5-minute aggregated data. The results are mentioned for an average run of each method. During the comparison, the parameters $u$ and $v$ for bagging are set to 70 and 400, respectively. While the optimal values of these parameters are subject to further discussion, they are chosen arbitrarily after multiple simulations. \par

\textcolor{black}{A further comparison between CQR and CQRb across multiple traffic measurement stations and various quantiles is illustrated on Figure \ref{fig:147_bagging} (additional examples are provided in Appendix A of the online supplement). In this comparison the data are initially aggregated by lane, not by road direction. The estimated CQRb curves are virtually indistinguishable from CQR ones. Otherwise, it would be always possible to eliminate the differences between them by increasing the number of grid points. In the case of the median ($\tau=0.5$) for TMS 147 illustrated on Figure \ref{fig:147_bagging} it was not possible to estimate using the default number of iterations. Therefore, it would be necessary to either increase the number of iterations, stretching the computational time, or use CQRb.} \par

\begin{figure}[ht]
    \centering
    \includegraphics[width=\textwidth]{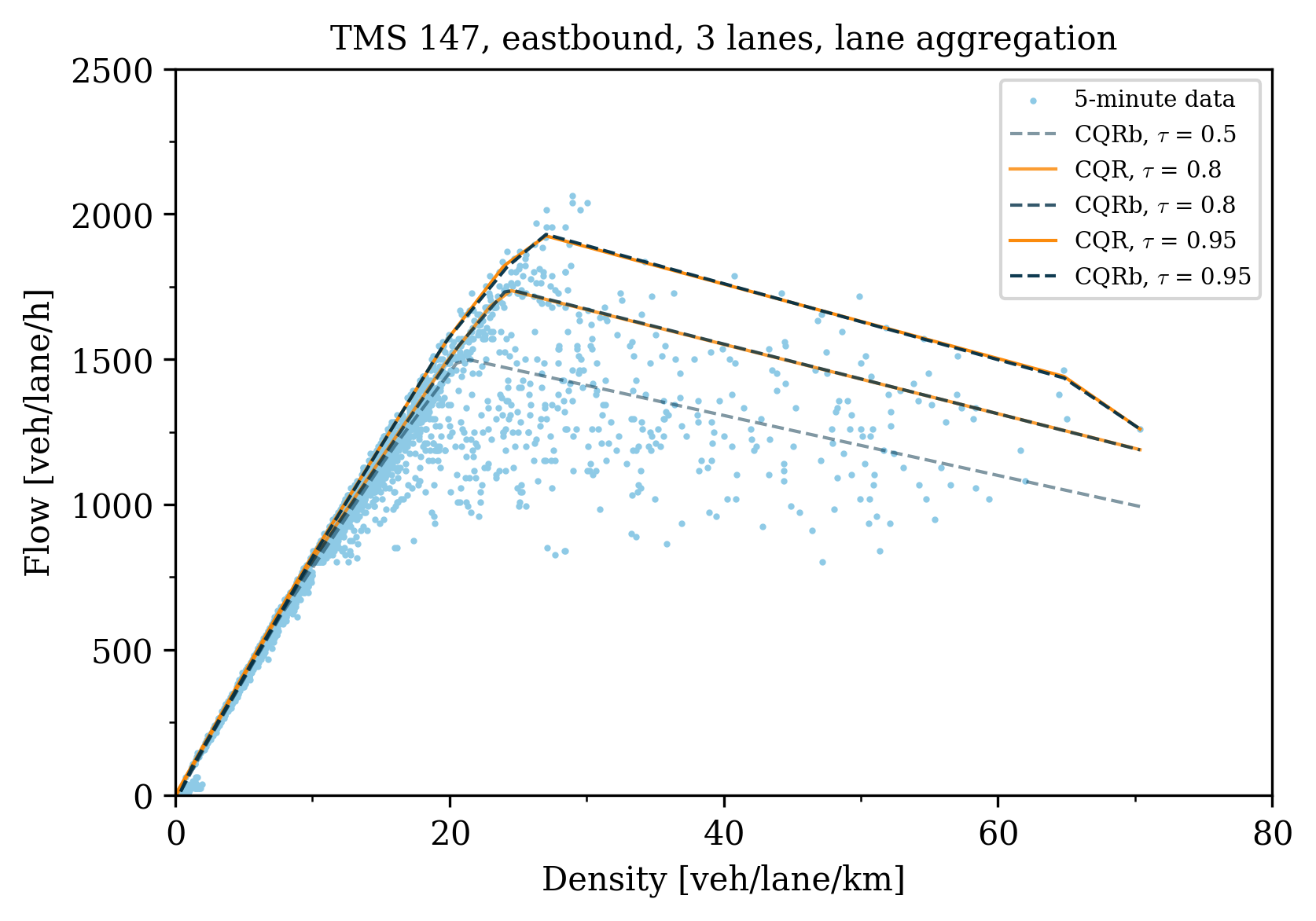}
    \caption{Comparison of CQR and CQRb across several quantiles}
    \label{fig:147_bagging}
\end{figure}

\textcolor{black}{In addition to reducing computational complexity, the bagging approach can provide other benefits. Most importantly, bagging can alleviate overfitting, as the approach plays the role of a regularization process (see \citet{hannah_multivariate_2013} for further discussion). In-depth assessment of bagging as a regularization method is out of scope of the current study and will be explored in more details in future research.}\par

\subsection{\textcolor{black}{Comparison of density-flow curves based on nearly stationary and time-aggregated data}} \label{sec:nearly_stationary}

\textcolor{black}{There are two primary approaches to data sampling when estimating the density-flow curve – using nearly stationary data and using time aggregated data \citep{nagel_still_2003, celikoglu_reconstructing_2013}. The first approach utilizes solely data points that represent nearly stationary conditions. This approach requires preliminary noise-reduction data cleaning, that is, removing the data points representing the non-stationary conditions, which results in the decreased scatter on the density-flow plane. Such noise reduction can help compensate for the limitations of the conventional deterministic approaches that rely heavily on the parametric functional form assumptions. To obtain nearly stationary data, it is necessary to have the data on either individual vehicles or aggregated over small time intervals (15-30 seconds). However, this is not always the case, and often data are initially provided in aggregated form with defined time intervals (e.g., 5 minutes). For international comparisons the Urban Traffic Data 2019 dataset by \cite{loder_understanding_2019} contains only aggregated data, and the data suppliers do not share the raw data. Hence, the alternative approach exists, which involves the use of time-aggregated data. Because road sensor data from Finland is provided on the level of individual vehicles, we will demonstrate the application of the convex quantile regression method on both nearly stationary and time-aggregated data.} \par

\textcolor{black}{To obtain data from nearly stationary states, the raw data for TMS 146 in the eastbound direction for the period from Monday, October 8, 2018 to Friday, October 12, 2018 is used. The data are divided into 30-second intervals and aggregated by the road direction, not separate lanes. Using these aggregated intervals the partitioning technique is applied. The resulted nearly stationary points are shown on Figure \ref{fig:near-stationary-single-0.95} with orange crosses. More figures could be found in Appendix B of supplementary material. As the number of data points is low, the CQR method is applied.} \par

\begin{figure}[ht]
    \centering
    \includegraphics[width=\textwidth]{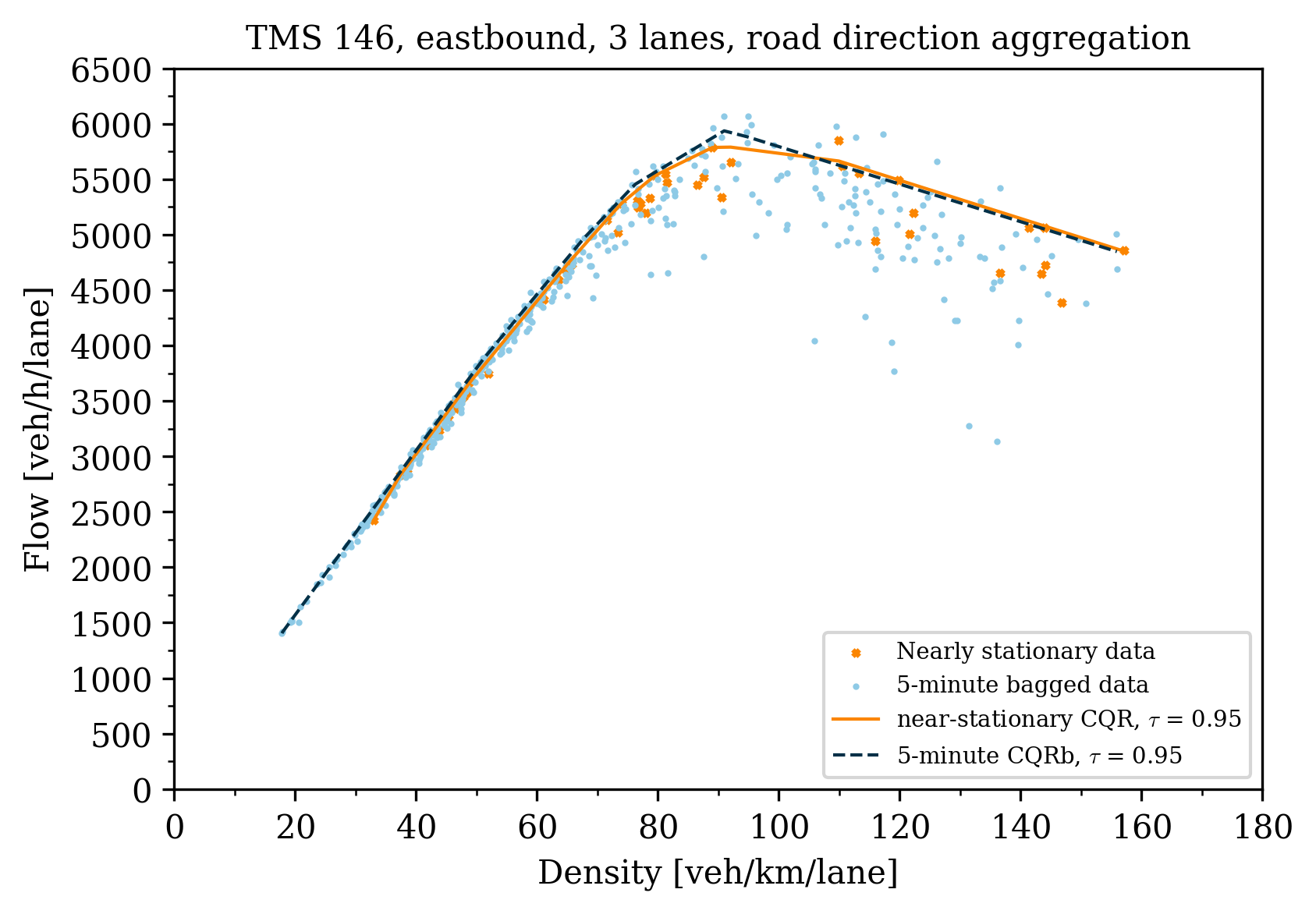}
    \caption{Comparison of estimates on nearly stationary and time-aggregated data}
    \label{fig:near-stationary-single-0.95}
\end{figure}

\textcolor{black}{For comparison, the data for the same period is aggregated into 5-minute intervals (time-aggregated data). This data has much more noise compared to the nearly stationary one. The curve is estimated using CQRb. Figure \ref{fig:near-stationary-single-0.95} demonstrates that curve estimated using time-aggregated data almost fully resembles the one estimated on nearly stationary data with $\tau=0.95$. The main difference occurs in the transition from the uncongested to the congested part of the diagram, because there are no observations representing nearly stationary conditions. The difference in capacity is less than 150 vehicles per hour.} \par 

\textcolor{black}{The example shows that the estimation using nearly stationary and time-aggregated data yields similar results, as convex quantile regression is less sensitive to noise compared to the ones with parametric functional form assumptions. As partitioning for nearly stationary data is resource expensive, we will continue with the application of the proposed method on time-aggregated data, what also broadens the scope on use to the datasets that provide data in aggregate format only. We leave more thorough examination of the convex quantile regression method in nearly stationary setting as an interesting avenue for future research.} \par

\subsection{Large-scale example based on time-aggregated data} \label{sec:time_agg}

In order to illustrate an example of the CQRb approach with estimation of critical density, data for TMS 146 for the whole year 2018 is used. The traffic measurement station is located on a road section featuring three eastbound lanes. To account for the specifics of the road traffic in Finland solely peak hours between 6:00 and 20:00 are analyzed. The original dataset includes more than 16 million observations of individual vehicles. These are aggregated into 5-minutes intervals, \textcolor{black}{over which harmonic average speeds are estimated}, for each lane separately, resulting in a total of $183\,465$ aggregated data points. Further we apply the bagging procedure with the density-axis divided into 20 equal segments, and the flow-axis divided into 200 equal segments. The final dataset includes $1\,614$ bags, which are used in the estimation. The number of bags is reduced compared to previous simulations. In our experience, we can have reasonable estimation time if the number of points is in range $1\,500\,-\,2\,000$. \par

\begin{figure}[ht]
    \centering
    \includegraphics[width=\textwidth]{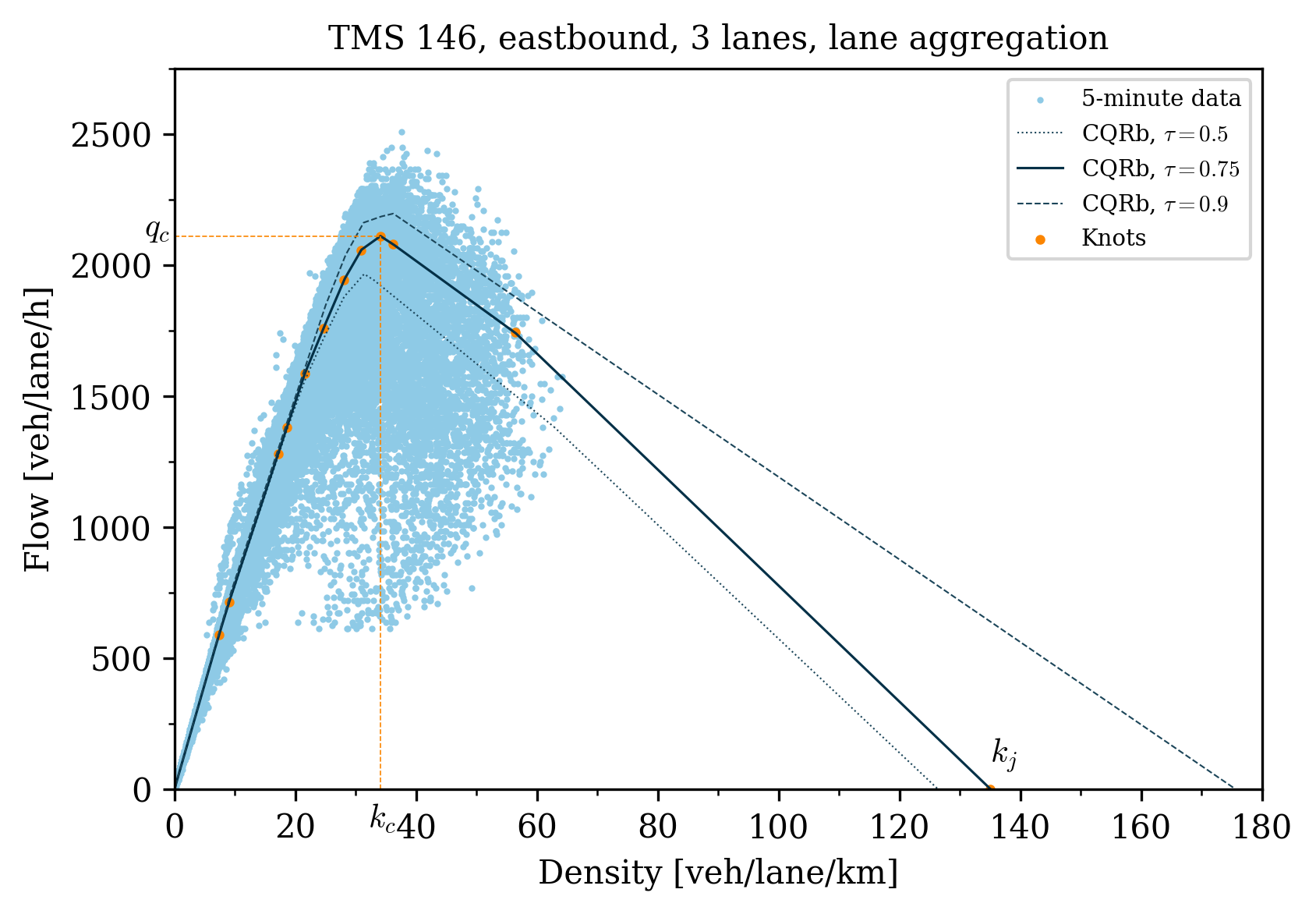}
    \caption{Density-flow curve estimation using CQRb}
    \label{fig:ill_exmp}
\end{figure}

Figure \ref{fig:ill_exmp} depicts the CQRb estimation using $\tau = 0.75$ on the aforementioned dataset, as well as $\tau = 0.5$ and $\tau = 0.9$. Following \citet{elefteriadou_revisiting_2006} here we discuss the case of $\tau = 0.75$ quantile which describes conditions that are frequently encountered. The estimation of a piecewise linear function has formed $12$ linear segments, $9$ of those located in the free-flow part and $3$ in the congested part of the \textcolor{black}{density-flow} diagram.  It is important to recall that the number of segments is not predetermined, but rather derived solely from the dataset. The equations for each piece are in (\ref{eq:CQRb_estimated_f}). 

The key characteristics of \textcolor{black}{traffic flow are easy to obtain from the estimation}. Capacity, which is taken as the maximum flow reached, is $q_c = 2109 \, vplph$ at the critical density of $k_c = 34 \, vplpkm$. The jam density is $k_j = 135 \, vplpkm$, which is obtained as $\frac{-\hat{\alpha}_{\max \mathrm{k}}^{bag}}{\hat{\upbeta}_{\max \mathrm{k}}^{bag}}$. It is important to note that despite using an exact number for the capacity, the CQRb method is still regarded as a stochastic capacity approach, as capacity varies depending on the selected quantile. There are 12 different wave speeds, each corresponding to a distinct linear segment with its own slope representing the shockwave speed. For the free flow part it is necessary to mention the decline of speed with the density growth, and the closer to the critical density, the faster the speed declines, which is depicted by the increasing number of line segments and their lower slope. This phenomenon aligns with the actual road conditions, as more vehicles on the road result in slower speeds and fewer opportunities for lane changes. The speed limit at the observed road section is $80 \, km/h$, and vehicles initially travel at this speed at low densities, slowing down to $70 \, km/h$ at the density of $26 \, vplpkm$ and flow of $1850 \, vplph$. At capacity the speed drops to $62 \, km/h$. The observed reduction in speed is substantial, however, it cannot be captured in the triangular diagram with the DGKHV calibration approach, as free-flow speed is assumed to be constant across all densities. \par

\begin{align}\label{eq:CQRb_estimated_f}
\scriptsize
\hat{q}(k)=
\begin{cases}
    79.95k            & 0       \leq k\leq 7.35 \\
    75.45k + 33.10   & 7.35   \leq k\leq 9.02 \\
    69.79k + 84.14  & 9.02   \leq k\leq 17.13 \\
    69.78k + 84.37   & 17.13   \leq k\leq 18.57 \\
    68.67k + 104.97 & 18.57   \leq k\leq 21.54 \\
    56.11k + 375.51 & 21.54   \leq k\leq 24.63 \\
    53.93k + 429.21 & 24.63   \leq k\leq 28.06 \\
    40.44k + 807.74 & 28.06   \leq k\leq 30.89 \\
    16.61k + 1543.74 & 30.89   \leq k\leq 34.02 \\
    -14.03k + 2585.99 & 34.02   \leq k\leq 36.14 \\
    -16.71k + 2683.16 & 36.14   \leq k\leq 56.33 \\
    -22.14k + 2988.56 & 56.33   \leq k\leq 135.01 \\
\end{cases}
\end{align}%

In the congested part there are three linear segments with wave speeds of  $-14.93, -16.72$ and $-22.14$. As congestion increases, the shockwave speed also increases in absolute value \textcolor{black}{due to the concavity constraints.}. The use of several wave speeds provides a more precise description of traffic behavior, demonstrating varying patterns as traffic saturation increases. Although we do not observe data points close to the jam density, we can still achieve meaningful estimations of this parameter.\par
\subsection{\textcolor{black}{Comparison of TMSs using separate lane aggregation}} \label{sec:overview}
\textcolor{black}{To follow the illustrative example, the results of comprehensive estimations are presented. To construct those estimations the following approach is used: for every TMS observed data from 2017 and 2018 is divided into consecutive two-week and one-month periods. For each of this periods the data are aggregated into 5-minute intervals by lane and by road direction. Then the CQRb model is estimated across 4 types of aggregations (road-week, road-month, lane-week, lane-month) for a set of $\tau \in \{0.5, 0.7, 0.75, 0.8, 0.85, 0.9, 0.95\}$ for 2 directions except TMS 109. Out of these estimations only the ones containing at least 15\% of bagged data points in the congested section ($k > \hat{k_c}$) are further considered to make sure that the congested section is represented. This results in 1346 estimations on lane-based aggregations and 4302 estimations on road-based aggregations included in the analysis.} \par

\begin{table}[]
\scriptsize
\centering
\caption{Results for separate lane aggregation}
\label{tab:average-results}
\resizebox{\textwidth}{!}{%
\begin{tabular}{@{}crrrrrrr@{}}
\toprule
TMS ID & \multicolumn{1}{c}{\begin{tabular}[c]{@{}c@{}}Average \\ capacity\\ {[}veh/hour/lane{]}\\ (st. dev.)\end{tabular}} & \multicolumn{1}{c}{\begin{tabular}[c]{@{}c@{}}Average \\ critical density \\ {[}veh/km/lane{]}\\ (st. dev.)\end{tabular}} & \multicolumn{1}{c}{\begin{tabular}[c]{@{}c@{}}Average \\ jam density\\ {[}veh/km/lane{]}\\ (st. dev.)\end{tabular}} & \multicolumn{1}{c}{\begin{tabular}[c]{@{}c@{}}Average \\ free flow speed \\ {[}km/h{]}\\ (st. dev.)\end{tabular}} & \multicolumn{1}{c}{\begin{tabular}[c]{@{}c@{}}Speed \\ limit \\ {[}km/h{]}\end{tabular}} & \multicolumn{1}{c}{\begin{tabular}[c]{@{}c@{}}Average \\ number of pieces\\ (st. dev.)\end{tabular}} & \multicolumn{1}{c}{\begin{tabular}[c]{@{}c@{}}Estimations \end{tabular}} \\ \midrule
118    & \begin{tabular}[c]{@{}r@{}}1381.06 \\ (170.12)\end{tabular}                                                        & \begin{tabular}[c]{@{}r@{}}29.6 \\ (5.77)\end{tabular}                                                           & \begin{tabular}[c]{@{}r@{}}127.45 \\ (13.11)\end{tabular}                                                           & \begin{tabular}[c]{@{}r@{}}59.36 \\ (5.18)\end{tabular}                                                              & 60                                                                                   & \begin{tabular}[c]{@{}r@{}}12.42 \\ (2.8)\end{tabular}                                                                                                                 & 28                \\ \midrule
10     & \begin{tabular}[c]{@{}r@{}}1561.92 \\ (95.42)\end{tabular}                                                         & \begin{tabular}[c]{@{}r@{}}30.11 \\ (7.42)\end{tabular}                                                          & \begin{tabular}[c]{@{}r@{}}128.51 \\ (10.1)\end{tabular}                                                            & \begin{tabular}[c]{@{}r@{}}66.75 \\ (7.63)\end{tabular}                                                              & 60                                                                                   & \begin{tabular}[c]{@{}r@{}}12.11 \\ (2.5)\end{tabular}                                                                                                                 & 36               \\ \midrule
116    & \begin{tabular}[c]{@{}r@{}}1720.64 \\ (105.77)\end{tabular}                                                        & \begin{tabular}[c]{@{}r@{}}28.76 \\ (3.1)\end{tabular}                                                           & \begin{tabular}[c]{@{}r@{}}137.58 \\ (11.06)\end{tabular}                                                           & \begin{tabular}[c]{@{}r@{}}71.67 \\ (4.89)\end{tabular}                                                              & 70                                                                                   & \begin{tabular}[c]{@{}r@{}}11.95 \\ (2)\end{tabular}                                                                                                                    & 70               \\ \midrule
145    & \begin{tabular}[c]{@{}r@{}}1848.94 \\ (197.3)\end{tabular}                                                         & \begin{tabular}[c]{@{}r@{}}29.15 \\ (5.62)\end{tabular}                                                          & \begin{tabular}[c]{@{}r@{}}133.91 \\ (13.18)\end{tabular}                                                           & \begin{tabular}[c]{@{}r@{}}81.97 \\ (6.06)\end{tabular}                                                              & 80                                                                                   & \begin{tabular}[c]{@{}r@{}}11 \\ (2.31)\end{tabular}                                                                                                                     & 209             \\ \midrule
146    & \begin{tabular}[c]{@{}r@{}}2017.67 \\ (205.26)\end{tabular}                                                        & \begin{tabular}[c]{@{}r@{}}31.39 \\ (5.19)\end{tabular}                                                          & \begin{tabular}[c]{@{}r@{}}133.76 \\ (12.53)\end{tabular}                                                           & \begin{tabular}[c]{@{}r@{}}75.38 \\ (6.43)\end{tabular}                                                              & 80                                                                                   & \begin{tabular}[c]{@{}r@{}}12.34 \\ (2.36)\end{tabular}                                                                                                                  & 214             \\ \midrule
147    & \begin{tabular}[c]{@{}r@{}}1774.78 \\ (170.49)\end{tabular}                                                        & \begin{tabular}[c]{@{}r@{}}24.63 \\ (3.08)\end{tabular}                                                          & \begin{tabular}[c]{@{}r@{}}136.65 \\ (12.04)\end{tabular}                                                           & \begin{tabular}[c]{@{}r@{}}83.73 \\ (8.93)\end{tabular}                                                              & 80                                                                                   & \begin{tabular}[c]{@{}r@{}}10.3 \\ (1.87)\end{tabular}                                                                                                                   & 185             \\ \midrule
148    & \begin{tabular}[c]{@{}r@{}}1856.86 \\ (154.73)\end{tabular}                                                        & \begin{tabular}[c]{@{}r@{}}25.25 \\ (3.05)\end{tabular}                                                          & \begin{tabular}[c]{@{}r@{}}134.97 \\ (11.83)\end{tabular}                                                           & \begin{tabular}[c]{@{}r@{}}87.21 \\ (4.73)\end{tabular}                                                              & 80                                                                                   & \begin{tabular}[c]{@{}r@{}}11.16 \\ (2.24)\end{tabular}                                                                                                                  & 314             \\ \midrule
149    & \begin{tabular}[c]{@{}r@{}}1802.67 \\ (139.87)\end{tabular}                                                        & \begin{tabular}[c]{@{}r@{}}26.8 \\ (2.64)\end{tabular}                                                           & \begin{tabular}[c]{@{}r@{}}129.99 \\ (11.59)\end{tabular}                                                           & \begin{tabular}[c]{@{}r@{}}81.47 \\ (6.4)\end{tabular}                                                               & 80                                                                                   & \begin{tabular}[c]{@{}r@{}}11.01 \\ (2.54)\end{tabular}                                                                                                                   & 129            \\ \midrule
11     & \begin{tabular}[c]{@{}r@{}}1303.14 \\ (40.69)\end{tabular}                                                         & \begin{tabular}[c]{@{}r@{}}25.96 \\ (2.4)\end{tabular}                                                           & \begin{tabular}[c]{@{}r@{}}134.71 \\ (11.43)\end{tabular}                                                           & \begin{tabular}[c]{@{}r@{}}63.24 \\ (4.99)\end{tabular}                                                              & 60                                                                                   & \begin{tabular}[c]{@{}r@{}}14.94 \\ (2.21)\end{tabular}                                                                                                                   & 54             \\ \midrule
109    & \begin{tabular}[c]{@{}r@{}}1998.2 \\ (179.22)\end{tabular}                                                         & \begin{tabular}[c]{@{}r@{}}22.99 \\ (2.25)\end{tabular}                                                          & \begin{tabular}[c]{@{}r@{}}132.11 \\ (12.11)\end{tabular}                                                           & \begin{tabular}[c]{@{}r@{}}108.53 \\ (6.14)\end{tabular}                                                             & 100                                                                                  & \begin{tabular}[c]{@{}r@{}}11.2 \\ (2.53)\end{tabular}                                                                                                                     & 107           \\ \bottomrule
\end{tabular}%
}
\end{table}

\textcolor{black}{Table \ref{tab:average-results} provides the comprehensive overview of the lane-based estimations. The table reports average capacity, density and jam density per lane as well as the average free flow speed (the $\upbeta$ of the first piece) and the average number of pieces that are received from the estimation. The number in brackets shows the standard deviation, which is rather high for the average capacity value. However, this behavior is expected, as results are averaged across different quantiles, and the capacity for $\tau=0.5$ is much smaller compared to the one for $\tau=0.9$. Capacity also varies between traffic measurement stations, as daily traffic differs between locations as shown in Table \ref{tab:data-info}.  What is necessary to point out is that the standard deviation for the jam density is higher than the one for the critical density. This happens due to the fact that the jam density is an extrapolation based on the $\hat{\upbeta}$ and $\hat{\alpha}$ of the last estimated piece. Therefore, when no heavy congestion is present, the extrapolation is rather long. Combined with varying quantiles this gives the described effect. Nevertheless, the average results of the estimations are in line with the generally observed. The results for the road direction aggregation are presented in Appendix C of supplement materials. More figures showing estimation examples could be found in Appendix D of supplement materials. }

\section{Out-of-sample \textcolor{black}{predictive power}} \label{sec:performance}
\textcolor{black}{In this section we compare the performance of the \textit{convex quantile regression with bags} (CQRb) approach with a calibration method for a triangular fundamental diagram introduced by \citet{dervisoglu_automatic_2009}, hereinafter DGKHV. In order to compare the out-of-sample predictive power of the methods, we partition the observed sample to two subsets referred to as the training set and the test set. The training set is used to estimate the model, specifically, the shockwave coefficients beta. The test set is not used for estimation, it is only utilized to evaluate the prediction accuracy of the two approaches. In-sample performance assesses the prediction accuracy of an estimator on the training dataset, while out-of-sample performance evaluates the estimator accuracy on a test dataset \citep{inoue_-sample_2005}. Although out-of-sample performance assessment is standard in the data science literature, to our knowledge, this study is the first out-of-sample performance assessment in the present context of the density-flow curve.} \par

The comparison of in-sample and out-of-sample performance incorporates the use of both mean absolute error ($\text{MAE} = \sum_{j=1}^{\vert U \times V \vert} \vert\hat{q}_i^{bag} - q_i^{bag}\vert$) and root mean squared error ($\text{RMSE} = \sqrt{\frac{1}{\vert U \times V \vert}\sum_{j=1}^{\vert U \times V \vert} (\hat{q}_i^{bag} - q_i^{bag})^2}$) as different loss functions are minimized in different methods \citet{geisser_predictive_2019} with errors being the lower the better. In terms of loss functions, for the DGKHV approach RMSE is expected to be lower compared to the proposed approach, whereas opposed holds for MAE. For the CQRb approach the MAE is expected to be theoretically lower compared to the DGKHV approach. \par

In every year data are grouped either by week, either by month periods. While aggregating data within each period into 5-minute time intervals, we consider two different approaches – aggregations by lane or by road direction. Aggregations by lane are aligned with \citet{dervisoglu_automatic_2009} and help describe  road sections near intersection where the behavior of traffic may vary based on lanes, while aggregations by road provide a better outlook for TMS located on road sections without nearby intersections. Thereby, there are 4 different aggregations applied to the same data: month-lane, month-road, week-lane, and week-road. Different types of aggregations are necessary to compare behavior between proposed CQRb approach and DGKHV calibration method. \par

The comparison between the proposed and the DGKHV approach is conducted through the following procedure. At first, data for one of the 4 aforementioned aggregations is taken for a selected period of time in 2016 or 2017, what forms the training set. Then, using this data, the \textcolor{black}{density-flow curve} is estimated using both the CQRb method and the DGKHV approach. For the CQRb method three different quantiles are estimated: $\tau \in \{0.75, 0.80, 0.85\}$. This helps us examine the differences between quantiles and account for any external factors that might still affect traffic. The estimated density-flow relationship will reveal the usual operating conditions of the road section. \par

Initially we compare the in-sample performance of two approaches to assess their ability to capture the variation in data. To make this comparison, 11210 test datasets are generated following the principles outlined above. On each dataset we conduct estimation using both CQRb and DGKHV approaches. Notably, the CQRb is estimated on the bagged data and subsequently extrapolated back with the use of the representation theorem to the same aggregated data thereby enabling the computation of errors on identical data. This way the comparison accurately reflects the differences between the two approaches. \par

We next examine the out-of-sample performance of the estimators, which assesses their ability to predict the density-flow relationship one year ahead at the same road section. \textcolor{black}{It is important to note that convex regression techniques such as CQRb (also CNLS, CQR) are guaranteed to satisfy global concavity both in-sample and out-of-sample, which is not the case for other nonparametric methods.} To account for fluctuating road conditions and traffic seasonality in Finland, we consider comparable time periods for out-of-sample performance. Specifically, for weekly data, we evaluate the same week of the following year compared to the in-sample period, while for monthly data, we assess the same month of the following year relative to the in-sample period, what forms the test set. For instance, if the estimation is done using March 2016 data, we evaluate prediction accuracy on March 2017 data. \par

Table \ref{tab:is_oos_errors} provides the averages for four distinct types of aggregations as well as the overall average, with error units measured in \textit{veh/h} for road aggregations and \textit{veh/lane/h} for lane aggregations respectively. Consequently it is expected to have higher errors in road aggregations set side by side with lane aggregations. In terms of \textcolor{black}{in-sample} performance, CQRb approach exhibits on average 40\% improvement in terms of MAE, which is anticipated as the quantile loss function minimizes absolute values. Furthermore, the in-sample RMSE of CQRb approach is smaller by 30\% on average, as well as for every type of aggregation. This error comparison provides into how well the estimates capture observed data, with the CQRb consistently delivering more precise results, as evidenced by the relative difference within the same type of error. \par

Out-of-sample predictions errors demonstrate that CQRb has on average a 6\% lower RMSE and 30\% lower MAE compared to the DGKHV approach. The DGKHV approach exhibits a 5\% lower RMSE for the week-lane aggregation, but still higher on average. \par

\begin{table}[ht]
\centering
\caption{RMSE and MAE comparison}
\label{tab:is_oos_errors}
\resizebox{\textwidth}{!}{%
\begin{tabular}{@{}lrrrrrrrr@{}}
\toprule
 & \multicolumn{2}{c}{in-sample RMSE} & \multicolumn{2}{c}{in-sample MAE} & \multicolumn{2}{c}{out-of-sample RMSE} & \multicolumn{2}{c}{out-of-sample MAE} \\ \midrule
 & DGKHV            & CQRb            & DGKHV            & CQRb           & DGKHV    & \multicolumn{1}{c}{CQRb}    & DGKHV    & \multicolumn{1}{c}{CQRb}   \\ \cmidrule(l){2-9} 
week-lane  & 127.99 & 99.76  & 73.19  & 50.85 & 137.17 & 143.38 & 77.59  & 61.77  \\
week-road  & 259.54 & 179.06 & 168.01 & 91.30 & 288.02 & 274.91 & 183.92 & 128.26 \\
month-lane & 142.57 & 105.17 & 80.69  & 50.42 & 150.63 & 129.75 & 85.05  & 58.87  \\
month-road & 298.55 & 184.19 & 192.19 & 83.79 & 308.94 & 227.41 & 200.52 & 107.33 \\ \midrule
average    & 199.69 & 140.60 & 124.14 & 70.25 & 216.42 & 202.70 & 133.46 & 92.49  \\ \bottomrule
\end{tabular}%
}
\end{table}

\begin{figure}[h]
    \centering
    \includegraphics[width=\textwidth]{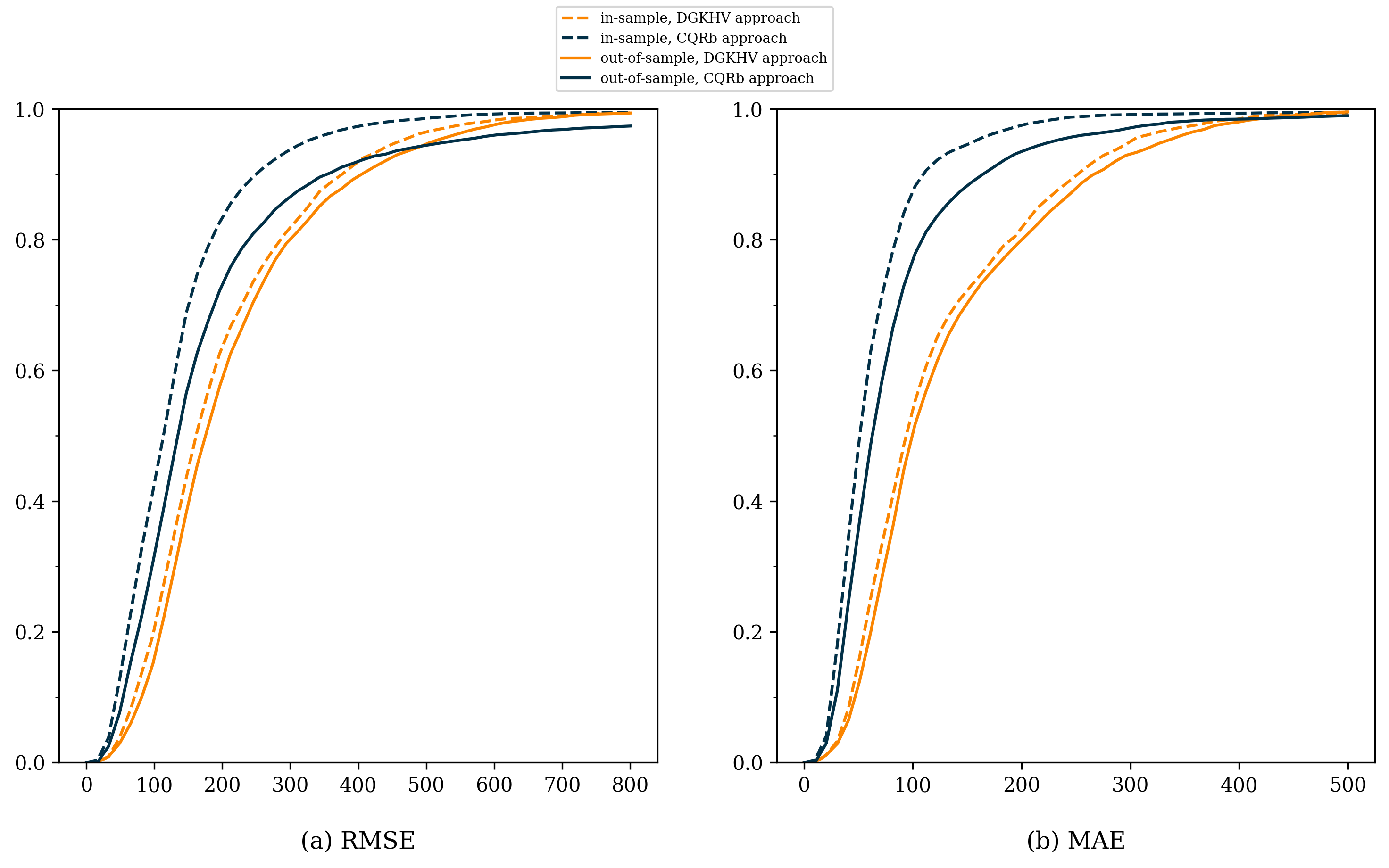}
    \caption{Cumulative distributions for RMSE and MAE}
    \label{fig:cum_dist_both_errors}
\end{figure}

The cumulative distributions of errors presented in Figure \ref{fig:cum_dist_both_errors} indicate that in case of in-sample errors the CQRb approach dominates the DGKHV approach in terms of the first-order stochastic dominance. In the out-of-sample case CQRb dominates the DGKHV approach in terms of the second-order stochastic dominance, with  92\% of RMSE errors and 99\% of MAE errors lower than the DGKHV approach, which further supports the superiority of the proposed approach. \par

Analyzing the prediction errors presents an opportunity to assess the predictive power of the proposed approach. \textcolor{black}{Recall that the out-of-sample performance statistics measure one-year ahead prediction accuracy.} Our findings indicate that the CQRb approach demonstrates lower root mean squared errors and mean absolute errors in both in-sample and out-of-sample performances, albeit by a smaller margin in the latter. Consequently, the proposed method is capable of predicting the density-flow relationship with greater precision. \par

The performance comparison highlights the remarkable performance of the introduced CQRb method, which consistently achieves lower errors both in-sample and out-of-sample, making it an efficient alternative to the DGKHV calibration method. The CQRb approach provides considerable flexibility in terms of understanding road data and responding to the changes of road state, without the need for deterministic assumptions. Different quantiles $\tau$ allow to describe levels of service in different conditions thereby offering a new perspective on traffic flow at a particular road section. With the use of quantiles the stochastic \textcolor{black}{relationships} are estimated, what is useful for the development and evaluation of control strategies \citep{ qu_stochastic_2017}. It is widely recognized that the performance of the highway systems can be substantially improved if the heterogeneity of traffic flow dynamics and the stochasticity of the \textcolor{black}{relationships} can be controlled \citep{ punzo_speed_2016}. \par

\section{Conclusions}\label{sec:conclusion}
\textcolor{black}{The objectives of this paper were threefold. Firstly,} we introduced a stochastic, fully nonparametric approach to estimating the \textcolor{black}{density-flow curve}. The method is an extension of the convex quantile regression (CQR) approach and belongs to the convex regression family of models. The piecewise linear form of the estimated function is capable of representing any arbitrary concave function, making it suitable for estimating the \textcolor{black}{density-flow curve} from empirical data, without having any restrictions on the number or location of pieces imposed. The proposed method satisfies the theoretical properties of the \textcolor{black}{density-flow curve} \textcolor{black}{and accounts for heterogeneity of traffic data through the estimation of multiple quantiles}.  \par

Secondly, we demonstrated how the methodology can be expressed in the form of a linear programming problem for a general case, and adapted it for the needs of traffic flow theory research \textcolor{black}{by introducing a new extension} termed \textit{convex quantile regression with bags}. The bagging approach \textcolor{black}{helps to alleviate} computational complexity of the method that arises from the large volume of traffic data necessary for an estimation. This approach substantially decreases the time required for the estimation of a regression function. \par 

The application of the proposed method to the real-world road sensor data from Finland shows how the classic properties of the \textcolor{black}{density-flow curve}  are followed with multiple linear segments describing both the free-flow and congested parts of the \textcolor{black}{density-flow} diagram offering a better capture of the variance in data. The linear nature of segments allows to determine the shockwave speeds and is preferential for traffic simulation purposes. \par

Finally, the paper demonstrated the predictive power of the CQRb approach through comparisons of mean absolute error (MAE) and root mean squared error (RMSE) in-sample and out-of-sample performance between CQRb and the DGKHV approach. Consistently lower errors across in-sample and out-of-sample estimations reveal the remarkable performance of the proposed approach and its ability to capture the variance in real-world data. \par

This study opens several interesting avenues for future research. For example, it would be worth to investigate how the results obtained from the estimation with the CQRb can be further used for road traffic simulation and modeling purposes. We believe that the introduced approach will provide researchers working with mesoscopic and macroscopic models a more data-driven perspective on the density-flow relationship. The approach can be highly beneficial for policy makers, as the new road control strategies can be developed from this more in-depth understanding of density-flow relationship. The \href{https://github.com/iarokr/roadtraffic}{\textit{roadtraffic}} Python package, which allows to apply the proposed method, is available on GitHub (\url{https://github.com/iarokr/roadtraffic}). \par

\section*{Acknowledgements}
The authors acknowledge the computational resources provided by the Aalto Science-IT project and the data provided by Fintraffic under the Creative Commons 4.0 BY license. Iaroslav Kriuchkov gratefully acknowledges financial support from Liikesivistysrahasto (210036) and KAUTE-säätiö (20220231, 20230211).

\section*{\textcolor{black}{Supplementary materials}}
\textcolor{black}{Supplementary material associated with this article can be found in the online version.}

{
\footnotesize
\setstretch{1.0}
\bibliography{references} 
}

\end{document}